# Rotterdam artery-vein segmentation (RAV) dataset

Jose D. Vargas Quiros[1,2], Bart Liefers[1,2], Karin A. van Garderen[1,2], Jeroen P. Vermeulen[1,2], EyeNED Reading Center[1,2], the VascX Research Consortium and Caroline C.W. Klaver[1,2,3,4]

[1]**Department of Ophthalmology, Erasmus University Medical Center, Rotterdam, the Netherlands**
[2]**Department of Epidemiology, Erasmus University Medical Center, Rotterdam, the Netherlands**
[3]**Department of Ophthalmology, Radboud University Medical Center, Nijmegen, the Netherlands**
[4]**Institute of Molecular and Clinical Ophthalmology, University of Basel, Switzerland**

## ABSTRACT

**Purpose:** To provide a diverse, high-quality dataset of color fundus images (CFIs) with detailed artery-vein (A/V) segmentation annotations, supporting the development and evaluation of machine learning algorithms for vascular analysis in ophthalmology.

**Methods:** CFIs were sampled from Dutch cohorts including the Rotterdam Study (RS), encompassing a wide range of ages, devices, and capture conditions. Images were annotated using a custom interface that allowed graders to label arteries, veins, and unknown vessels on separate layers, starting from an initial vessel segmentation mask. Connectivity was explicitly verified and corrected using connected component visualization tools.

**Results:** The dataset includes 1024x1024-pixel PNG images in three modalities: original RGB fundus images, contrast-enhanced versions, and RGB-encoded A/V masks. Image quality varied widely, including challenging samples typically excluded by automated quality assessment systems, but judged to contain valuable vascular information.

**Conclusion:** This dataset offers a rich and heterogeneous source of CFIs with high-quality segmentations. It supports robust benchmarking and training of machine learning models under real-world variability in image quality and acquisition settings.

**Translational Relevance:** By including connectivity-validated A/V masks and diverse image conditions, this dataset enables the development of clinically applicable, generalizable machine learning tools for retinal vascular analysis, potentially improving automated screening and diagnosis of systemic and ocular diseases.

# Introduction

The retinal vasculature offers a unique and non-invasive window into the human body's microcirculation. Morphological characteristics of retinal blood vessels, such as their caliber, tortuosity, and branching patterns, serve as important biomarkers for systemic health[1,2]. Deviations in vascular metrics have been robustly linked to a variety of systemic conditions, including hypertension, cardiovascular disease, and stroke[3–5]. More recently, these ocular biomarkers have also shown associations with neurodegenerative and psychiatric disorders, such as Alzheimer's disease and schizophrenia[6,7], further highlighting the potential of *oculomics* in modern medicine[8].

The accurate measurement of these biomarkers is predicated on the precise segmentation of the vascular tree from color fundus photographs, followed by the correct classification of each vessel segment as either an artery or a vein. Historically, this process was manual or semi-automated, making it labor-intensive and unsuitable for large-scale studies. The advent of deep learning has revolutionized this field, enabling the development of fully automated pipelines for vessel segmentation and classification[9–11]. However, the performance and generalizability of these data-driven models are fundamentally dependent on the availability of large, diverse, and meticulously annotated training datasets. While several datasets for artery-vein segmentation exist, such as RITE (AV-DRIVE)[12] and HRF[13], the number and size of datasets with specific artery-vein (A/V) labels remains limited.

Table 1 provides a summary of existing A/V segmentation datasets. While these datasets have been instrumental in advancing the field, many are constrained by their relatively small size, a lack of diversity in imaging devices and conditions, or a focus on very specific patient demographics (i.e., macula- or optic disc-centered). A need persists for larger scale datasets with a wider variety of imaging devices, a broad spectrum of common ocular diseases, and both macula- and disc-centered images to foster the development of more robust and generalizable algorithms.

To address these gaps, we introduce the Rotterdam Artery-Vein (RAV) dataset, a large and diverse collection of fundus images with detailed A/V annotations from the population-based Rotterdam Study[14]. The RAV dataset contains 206 retinal fundus images from well-characterized cohorts. All images and annotations are made publicly available under a CC license. The RAV dataset was used to train the models in VascX, a publicly available system that includes models for artery-vein segmentation. The system was shown to achieve state-of-the-art segmentation performance on various datasets[11].

**Table 1.** Publicly available datasets with artery-vein annotations. (H: Healthy, DR: Diabetic Retinopathy, GC: Glaucoma, NTG: Normal-Tension Glaucoma, HTG: High-Tension Glaucoma, O: Other, M: Macula-centered, OD: Optic Disc-centered, NR: Not Reported).

| Dataset | Images | Country | Age (years) | Pathologies | Center | FoV |
|---|---|---|---|---|---|---|
| DRIVE-AV[15] | 20 | NL | 25-90 | 7 DR, 13 no DR | M | 45 |
| RITE[12] | 40 | NL | 25-90 | 7 DR, 33 no DR | M | 45 |
| CHASE DB1[16] | 28 | UK | 11-13 | Healthy children | M | 30 |
| Les-AV[17] | 22 | NR | NR | NR | OD | |
| HRF-AV[18] | 45 | GER | NR | 15 H, 15 DR, 15 GC | M | 45 |
| Rotterdam (ours) | 206 | NL | 40+ | See Figure 1. | | |

The dataset is characterized by its high diversity, incorporating images from multiple camera systems, a wide range of common ocular pathologies, and a mixture of macula- and optic disc-centered fields of view.

# Methods

## Imaging sources

The dataset consists of images from Dutch cohorts available to the Department of Ophthalmology at Erasmus Medical Center in Rotterdam; mainly the Rotterdam Study (RS) . The RS is a prospective population-based cohort study of people living in Ommoord, a district of the city of Rotterdam. The RS consists of four cohorts, all of which were used in this work. The minimum age of study participants varies between >55 in the first cohort and >40 in the fourth. Each cohort was followed for multiple rounds of follow-up examinations every 4 to 5 years [14].

MYST was conducted from 2010 to 2012. Participants were recruited via public media, eye care clinicians, and announcements on websites from the study or from Erasmus Medical Center [19].

AMD-Life is a 2-year lifestyle intervention trial. The first year is a self-contained open-label randomized clinical trial with 3 intervention arms. The second year aims to investigate the sustainability of improved lifestyle without active intervention. Patients visit the research center 3 times: at baseline, month 12, and month 24.[20].

Most of the patient visits in these studies involved the capture of CFIs on one or both eyes. Due to the multi-decade span of the RS, multiple devices, capture conditions and fields (macula and disc centered) are present in the dataset.

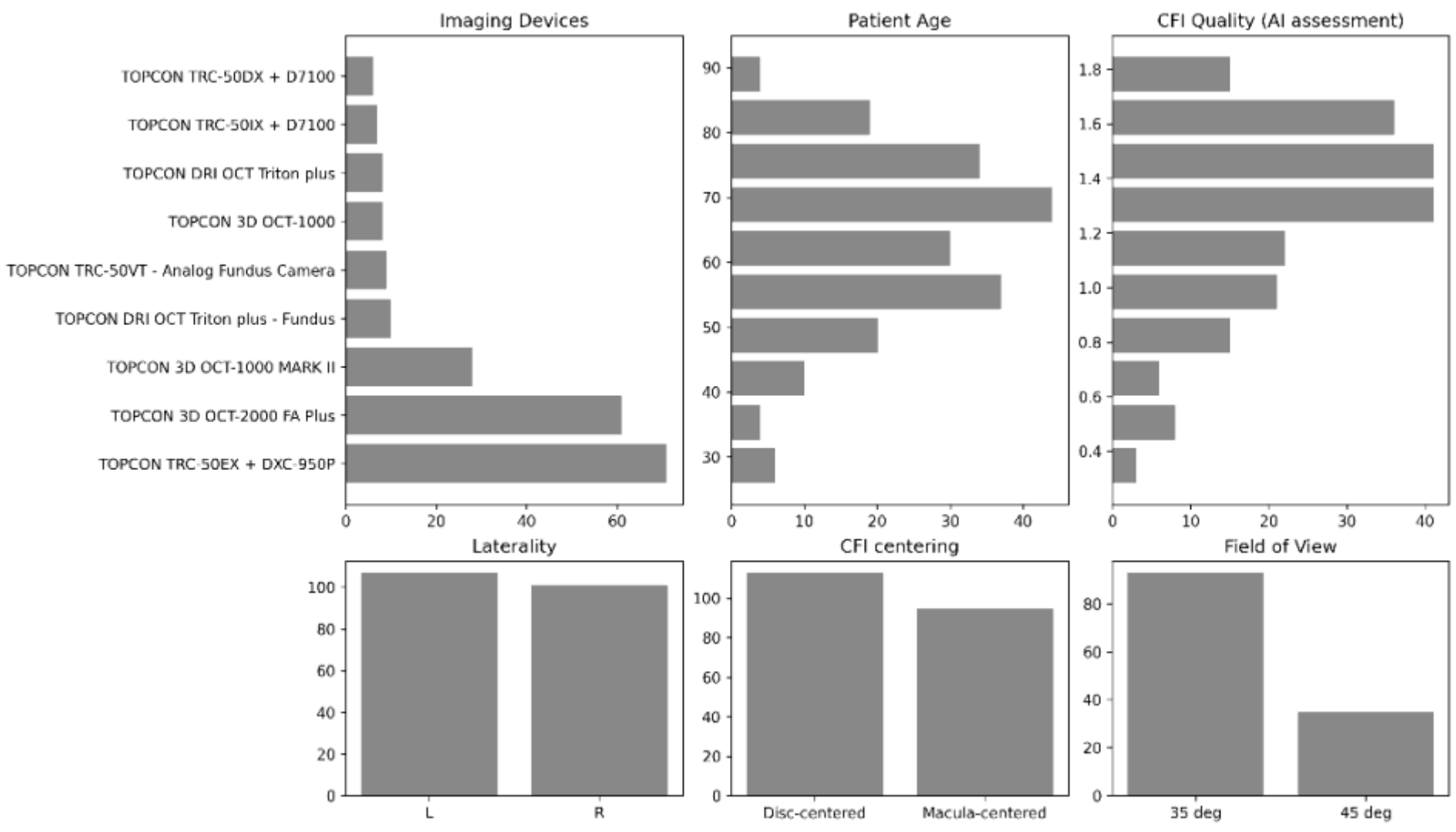

*Figure 1.General characteristics of the Rotterdam Artery-Vein Segmentation dataset CFIs (206 images)*

## Sampling and Quality Control

We sampled CFIs at random from all available images. We included fundus images captured by OCT machines and older analog devices to improve diversity of devices in the dataset. Due to the much larger size of the Rotterdam Study compared to AMD-Life and MYST, we over-sampled CFIs from these sets to make up roughly 15% of the samples. We asked graders to exclude images for which they found the quality to be completely unusable for artery-vein segmentation due to overall low visibility of the vessels. However, we were intentionally inclusive regarding quality to encourage a more diverse and challenging development set. We included images considered unusable by quality estimation algorithms, under the consideration that although the general quality level of these images is poor, many contained clearly usable vascular information for measuring certain biomarkers.

## Human Labelling of Arteries and Veins

Due to the high performance of existing vessel segmentation algorithms (which segment the entire vasculature without discrimination between arteries and veins), we designed our artery-vein labelling process around the coloring or labelling of vessel segmentation masks. Graders were provided with a high quality vessel segmentation output by a model and asked to *split* the vessel segmentation mask into separate *Artery* and *Vein* masks. In A/V crossings (where an artery crosses over a vein or viceversa) we labelled on both the artery and vein masks as opposed to labelling only the visible / superior vessel. This allows models trained on the segmentations to potentially resolve the topology of the underlying vascular trees.

An interface was developed specifically for artery-vein segmentation. The interface enables graders to color arteries and veins into two overlapping masks using drawing tools. A third mask - Unknown was added for vessels that could not be recognized as either. They were able to visualize and correct each mask independently, or all at the same time. A button in the interface  allowed them to visualize the connected components of each mask in different colors to more easily find mistakes in connectivity. The entire process included:

1. Correcting mistakes in the provided mask (vessel segmentation).
2. Splitting the mask into Artery, Vein and Unknown masks using drawing tools.
3. Verifying the connectivity of the masks and filling any clear gaps.

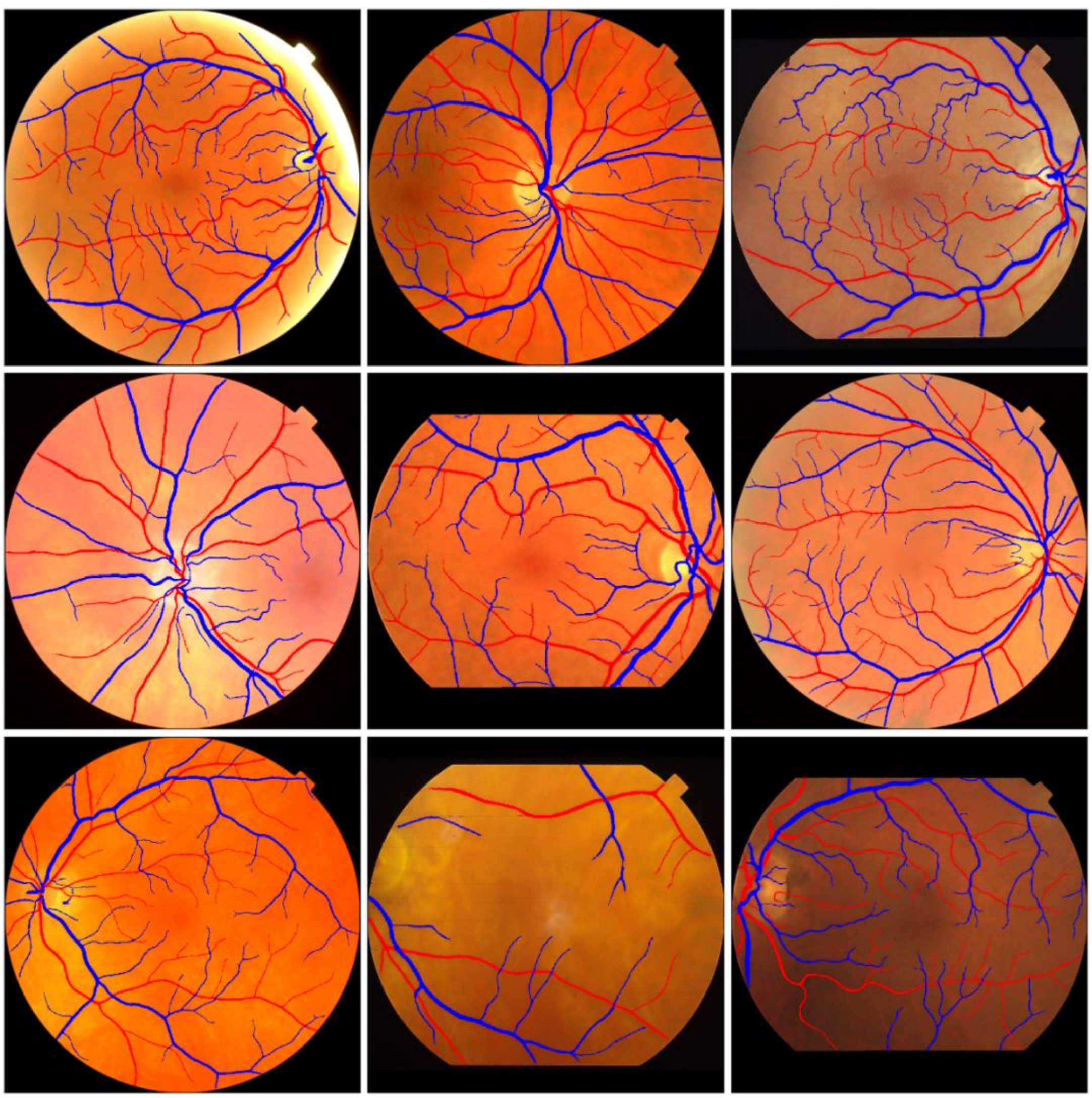

*Figure 2. Sample segmentations from the Rotterdam Artery-Vein segmentation dataset.*

# Data Summary

The data is arranged into three folders, all of which contain PNG files at 1024x1024 pixels with matching names.

**rgb**: preprocessed color fundus images in 1024x1024 resolution. The images were cropped to their circular bounds from their original dimensions for consistency and in some cases to remove identifiers present in the background.

**av_masks**: artery-vein segmentation masks. The masks are encoded in the RGB channels. R = arteries, B = veins, G = unknown

Metadata fields

A metadata file is included with basic information about the images:

- **device**: manufacturer and model
- **patient_age**: patient age in years at time of acquisition
- **patient_sex**: M / F

# Technical Validation

We measured inter-rater agreement on our A/V annotation process by asking four graders to label the same independent set of images. To reduce the workload and maintain a sufficient sample of images, we asked the graders to label only a (randomly positioned) window within each CFI, within which we measured agreement in the labelled artery and vein masks, separately. Agreement was measured using both Kappa and mean Dice score between pairs of graders.

*Table 1. Inter-rater agreement mean (standard deviation) between our four graders on artery and vein labelling.*

| | Cohen's Kappa | Dice Score |
|---|---|---|
| Arteries | 0.882 (0.146) | 0.906 (0.078) |
| Veins | 0.890 (0.106) | 0.899 (0.099) |

# Discussion

The RAV dataset is a contribution to the biomedical imaging community by supporting the development and evaluation of better artery-vein segmentation models. Improved segmentation will in turn result in improved calculation of downstream biomarkers. The size and variability in the dataset is likely to support the creation of more robust models that are more likely to generalize to different devices and conditions[11].

The methodology used to produce the RAV dataset is also an important contribution in enabling the time and cost-efficient annotation of a large CFI dataset, thanks to the initialization with AI-generated vessel segmentations. This allowed us to obtain high quality annotations on a large image set, with human verification to ensure its quality. Note that due to the nature of our labelling process, inter-rater agreement is likely over-estimated when compared to the agreement in a labelling process where images are labelled from scratch, without the aid of a model segmentation.

We hope that our work with the RAV dataset will encourage researchers to follow our example in publishing larger scale datasets across a wider variety of imaging and patient conditions; thereby supporting the development of the technical work that is necessary in the field.

## Data Repository

The RAV dataset is available at:

https://dataverse.nl/dataset.xhtml?persistentId=doi:10.34894/9OIMWY

All the data was made publicly available under a Creative Commons Attribution-NonCommercial 4.0 International license. Additional images may be available upon request for research purposes under a EULA (End License User Agreement).

All images and segmentations have been preprocessed to a standardized 1024x1024px resolution.

## Acknowledgments

The authors acknowledge the VascX consortium for supporting this project. The authors are grateful to the study participants and the staff from the Rotterdam Study.

This work was funded by the Swiss National Science Foundation grant no. CRSII5 209510. Supported by Oogfonds, Henkes stichting, and Landelijke Stichting voor Blinden en Slechtzienden (LSBS). Additional support was provided by Erasmus Medical Center, Erasmus University, Netherlands Organization for the Health Research and Development (ZonMw), and the Ministry of Education, Culture and Science,

# Author Information

The VascX Research Consortium (in alphabetical order):

Ciara Bergin[1]; Sven Bergmann[1,2,4]; Michael Beyeler[1,2,5]; Dennis Bontempi[1,2]; Sacha Bors [1,2]; Leah Böttger[1,2]; Bogdan Draganski[5,6,7]; Adham Elwakil[3,8]; Györgyi V. Hamvas[9,10]; Janna Hastings[11,12]; Ilaria Iuliani[1,2]; Caroline C.W. Klaver[13,14,15,16]; Ihor Kuras[6,7]; Bart Liefers[13,14]; Ilenia Meloni[3,8]; Sofia Ortin Vela[1,2]; David Presby[1,2]; Ian Quintas[1,2]; José D. Vargas Quiros[13,14]; Marc Schindewolf [9,10]; Reinier O. Schlingemann[3,17]; Mattia Tomasoni [3,8]; Olga Trofimova[1,2].

1 Department of Computational Biology, University of Lausanne, Lausanne, Switzerland

2 Swiss Institute of Bioinformatics, Lausanne, Switzerland

3 Department of Ophthalmology, University of Lausanne, Fondation Asile des Aveugles, Jules Gonin Eye Hospital, Lausanne, Switzerland.

4 Department of Integrative Biomedical Sciences, University of Cape Town, Cape Town, South Africa

5 Insel University Hospital Bern, Switzerland

6 Department of Neurology, Max Planck Institute for Human Cognitive and Brain Sciences, Germany

7 Department of Clinical Neuroscience, Lausanne University Hospital and University of Lausanne, Switzerland

8 Platform for Research in Ocular Imaging, Fondation Asile des Aveugles, Jules Gonin Eye Hospital, Lausanne, Switzerland.

9 Inselspital, Bern University Hospital, University of Bern, Switzerland.

10 Department for BioMedical Research, Bern University Hospital, University of Bern, Switzerland.

11 Institute for Implementation Science in Health Care, Faculty of Medicine, University of Zurich, Zürich, Switzerland

12 School of Medicine, University of St Gallen, St. Gallen, Switzerland

13 Department of Ophthalmology, Erasmus University Medical Center, Rotterdam, The Netherlands.

14 Department of Epidemiology, Erasmus University Medical Center, Rotterdam, The Netherlands.

15 Department of Ophthalmology, Radboud University Medical Center, Nijmegen, the Netherlands.

16 Institute of Molecular and Clinical Ophthalmology, University of Basel, Switzerland.

17 University Medical Centres, Amsterdam, The Netherlands.